\documentclass{aa} %
\usepackage{graphicx}
\usepackage{txfonts}
\begin{document}

\title{Secrets behind the $RXTE$/ASM light  curve  of  Cyg X-3}
\subtitle{ Feedback between wind-fed accretion and luminosity }

  \author{O. Vilhu
         \inst{1}
               \and
    K.I.I. Koljonen
       \inst{2}
\and
 D.C. Hannikainen
  \inst{3}
           }
 \offprints{O. Vilhu}
  \institute{Dept of Physics, P.O. Box 84, FI-00014  University of Helsinki, Finland\\
              \email{osmi.vilhu@gmail.com}
\and
       Institutt for Fysikk, Norwegian University of Science and Technology, H\"{o}gskoleringen 5, Trondheim, 7491  Norway\\
        \email{karri.koljonen@ntnu.no}
     \and
           Sky \& Telescope, 1374 Massachusetts Ave Floor 4, Cambridge, MA 02138, USA\\
                   \email{diana@skyandtelescope.org}
     }

   \date{Received ; accepted  }

  \abstract
{In wind-fed X-ray binaries, the radiatively driven wind of the primary star can be suppressed by the X-ray irradiation of the compact secondary star, leading to an increased accretion rate. This causes feedback between the released accretion power and the  luminosity of  the compact star (X-ray source).}
{ We investigate the feedback process between the released accretion power and the X-ray luminosity of the compact star (a low-mass black hole) in the unique high-mass X-ray binary Cygnus X-3. We study whether the seemingly erratic behavior of the observed X-ray light curve and accompanying spectral state transitions could be explained by this scenario. }
{ The wind-fed accretion power  is positively correlated with the extreme-ultraviolet (EUV) irradiation of the X-ray source. It is also larger than the  bolometric luminosity  of the X-ray source derived by spectral modeling  and assumed to be an intrinsic property of the source.  We assume that a part  of  the wind-fed power  experiences a small amplitude variability around the source luminosity. The largest luminosity (lowest wind velocity) is constrained  by the Roche-lobe radius, and the lowest one is constrained by the  accretion without EUV irradiation. There is a delay between the EUV flux  fixing the wind-fed power and that from the source. We modeled this feedback   assuming different time profiles for the small amplitude variability.}
{  We propose a simple heuristic model to couple the influence of EUV irradiation on the stellar wind (from the Wolf-Rayet companion star) with the X-ray source itself.   The resulting 
time profile of  luminosity mimics that of the input variability, albeit with a larger amplitude.  The most important property of the input variability are turnover times when it changes its sign and starts to have either positive or negative feedback. The bolometric luminosity derived by spectral modeling is the time average of the resulting feedback luminosity.
                                                                                 }
{      We demonstrate that the erratic behavior of the X-ray light curve of Cygnus X-3 may have its origin in  the small amplitude variability of the X-ray source and feedback with the companion wind. This variability could arise in the accretion flow and/or due to the loss of kinetic energy in a jet or an accretion disk wind. In order to produce similar properties of the simulated light curve as observed, we have to  restrict the largest accretion radius to a changing level, and assume variable timescales for the rise and decline phases of the light curve.
     }   

   \keywords{Stars:individual:Cyg X-3 -- X-rays:individual:Cyg X-3 – Stars:black holes -- Stars:binaries:close  }
      \maketitle        

\section{Introduction}

Cygnus X-3 (4U 2030+40), one of the first X-ray binaries  discovered (Giacconi et al. 1967), is also one of the brightest in the X-ray and radio regimes. It is listed in the catalogue of High-Mass X-ray Binaries (HMXB) in the Galaxy (Liu et al. 2006), where most entries consist of an OB star whose wind feeds a companion neutron star or black hole, releasing X-rays in the process.  The donor star of Cyg X-3 (optical counterpart V1521 Cyg) is a Wolf-Rayet (WR) star either of  type WN 5-7 (van Keerkwijk et al.  1996) or a weak-lined WN 4-6 (Koljonen \& Maccarone 2017). The relatively small size of this helium star allows for a tight orbit with a period of 4.8 hours which is more typical of low-mass X-ray binaries. Cyg X-3 is often classified as a microquasar and the radio, infrared, and X-ray properties  suggest that the companion star is a low-mass black hole, although a neutron star cannot be totally ruled out (Zdziarski et al. 2013).

Vilhu et al. (2021) show quantitatively how the wind velocity of the WR star on the face-on side depends on the extreme-ultraviolet (EUV) irradiation (at 100 eV) from the compact star, in agreement with the results by Krticka et al. (2018) for irradiation effects in HMXBs. Increasing the EUV leads to the suppression of wind velocity. This, in turn, increases the accretion rate and released accretion power due to the more effective wind capture rate of the compact star.     Therefore, there exists a positive correlation between the EUV irradiation and  the power provided by the wind accretion.

 A similar correlation exists between the EUV flux and the bolometric luminosity    $L_{bol}$  (= $L_{eqpair}$) released by the X-ray source itself as demonstrated by Hjalmarsdotter et al. (2009) using the hybrid thermal and nonthermal plasma emission model EQPAIR (Coppi 1992). 
The variability of the EUV irradiation leading to changes in the wind velocity and subsequently the wind-fed accretion luminosity may be  behind the evolution of the X-ray $RXTE$/ASM light curve of Cyg X-3 \footnote[1]{https://heasarc.gsfc.nasa.gov/docs/xte/asm-products.html}.  

A part of the wind-fed  power  goes to the observed radiation of the X-ray source ($L_{rad}$), while a part is advected to the black hole and/or lost from the system by jets or accretion disk winds.  In the present paper, we propose a heuristic model   by balancing between  $L_{rad}$ and  $L_{eqpair}$. These should be equal but we  assume a small variability around this equality; $L_{eqpair}$ is a time average of $L_{rad}$ and characteristic of the X-ray source.

\section{Methods}    
\begin{figure}
   \centering
   \includegraphics[width=9cm]{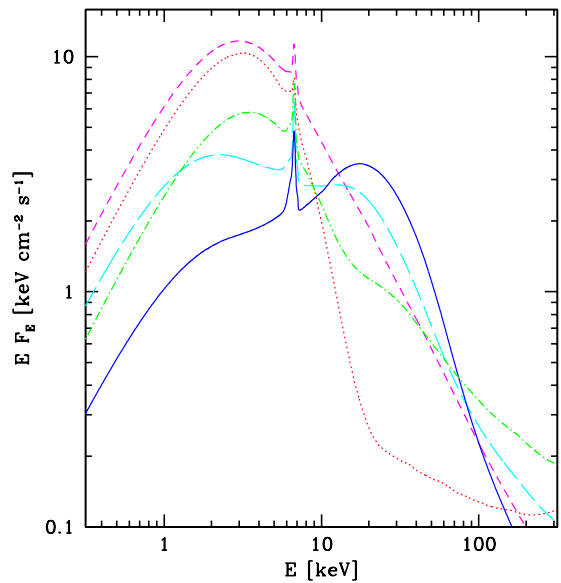}
     \caption{Unabsorbed model spectra of the five main spectral states of Cyg X-3 reproduced from Hjalmarsdotter et al. (2009). The models corresponding to  the hard state (solid blue line), the intermediate state (long-dashed cyan line), the very high state (short-dashed magenta line), the soft nonthermal state (dot-dashed green line), and the ultrasoft state (dotted red line) are shown (unabsorbed at Earth).}
         \label{spectralstates}
   \end{figure}

Hjalmarsdotter et al. (2009) analyzed an extensive set of $RXTE$ spectra of Cyg X-3 and fitted them with EQPAIR spectral modeling by Coppi (1992) during five spectral states ( see Fig.~\ref{spectralstates}). The bolometric luminosities   $L_{bol}$ (= $L_{eqpair}$) of these states are plotted in Fig.~\ref{LvsEUV} versus the EUV flux ($EF_E$ at 100 eV) at the system using a distance  of 7.4 kpc (McCollough et al. 2016) and correcting the fluxes given in Vilhu et al. (2021) to this distance (dashed line). The system is optically too weak for a Gaia distance estimate (due to strong absorption). We also corrected the fluxes  for orbital modulation using the observing log in Hjalmarsdotter et al. (2009),  estimating the modulation between 2 - 100 keV  from  Vilhu et al. (2003,  see their Fig. 3 and the folding ephemeris there). This amounted to a correction factor 1.3 - 1.4. The fluxes are given in Table 1.

For comparison, the wind-fed accretion power $P_{acc}$ is shown in Fig.~\ref{LvsEUV} (solid line) as computed from the Bondi-Hoyle-Lyttleton (BHL) formalism using a wind model with clumping volume filling factor 0.1 (see Eqs. 2 - 5 and Table 1). Clumpy winds are ubiquitous in hot stars. The value 0.1 means that all the wind mass in clumps comprises 10\% of the total wind volume. On average, the wind-fed accretion power is a factor of 2 larger than the bolometric luminosity likely arising from the power lost in the advective and/or kinetic components (vertical line in  Fig.~\ref{LvsEUV}). The wind-fed power without the effect of EUV irradiation would result in  1.35$\times$10$^{38}$  erg/s (horizontal line in  Fig.~\ref{LvsEUV}). 

We adopted  masses of 2.4 $M_{\odot}$ and 10$M_{\odot}$ for the compact star and WR star, respectively, and  the wind mass-loss rate  $\dot M_{wind}$ = 6.5$\times$10$^{-6}$  $M_{\odot}$/year used in Vilhu et al. (2021). These all are mean values within error bars given by Zdziarski et al. (2013). Their solution allows for the presence of either a neutron star or a low-mass black hole, but they consider that  the radio, infrared, and X-ray properties of the system suggest that the compact star is a black hole. In the present paper, we adopt this view.

Wind models were computed with the methods explained in Vilhu et al. (2021), integrating the mass conservation and momentum balance equations with the upward booster (line force). The computation of the line force is explained in Appendix B of Vilhu et al.  utilizing the XSTAR data-base \footnote[2]{https://heasarc.gsfc.nasa.gov/xstar/xstar.html}. The EUV irradiation weakens the booster and, hence, lowers the wind velocity.

\begin{figure}
\centering
\includegraphics[width=9cm]{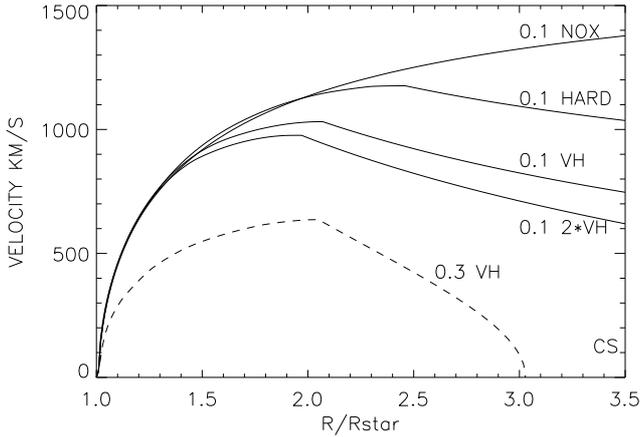}
   \caption{Wind  velocity models of the WR component at the face-on side are plotted versus  distance from the WR surface. The models were computed without the gravitational pull of the compact star (the distance of which  is marked by CS).   The numbers 0.1 and 0.3 along the curves  mean clumping volume filling factors $f_{vol}$. The labels HARD and VH mean the wind during the hard and very high states, respectively; and 2*VH means doubling of the very high state EUV flux. The kinks (discontinuities) are due to stopping the line force.}
    \label{loop}
 \end{figure}

\begin{figure}
   \centering
   \includegraphics[width=9cm]{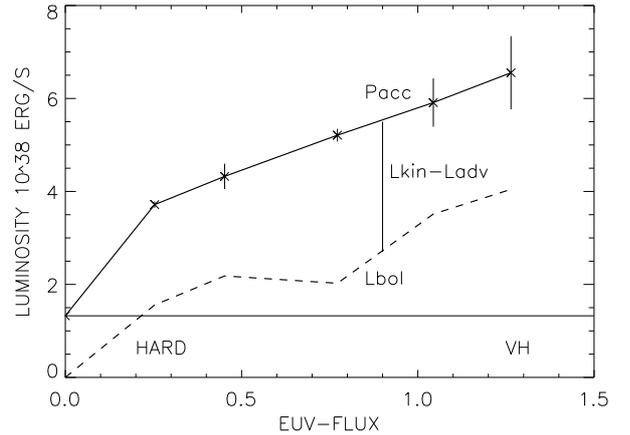}
     \caption{Wind-fed power $P_{acc}$  of Cyg X-3 as a function of the EUV irradiation ($EF_E$ in units of 2$\times$$10^{36}$ erg/s at 100 eV)  as computed with the Bondi-Lyttleton formalism  (solid line with error bars) from the wind velocities of Table 1. A similar relation for the bolometric luminosity $L_{eqpair}$ =$L_{bol}$ of the X-ray source itself is shown by the dashed line, and it was obtained by  EQPAIR modeling of X-ray spectra (Hjalmarsdotter et al. 2009). The horizontal line shows the  power provided by the wind without EUV irradiation (NOX). The positions of the hard and very high spectral states are marked (HARD and VH). The vertical line shows the amount of power going to kinetic and advective parts. }
         \label{LvsEUV}
\end{figure}

In the modeling we used a wind clumping volume filling factor $f_{vol}$ = 0.1 similar to that in Vilhu et al. (2021, Table A1, mean of phase angles $0^\circ$ and $30^\circ$), except that the nonirradiated model (NOX) has the velocity at infinity $V_{inf}$ = 1600 km/s (as in WNE-w stars, Hamann et al. 1995, their Fig. 3). For the different spectral states (no EUV irradiation, NOX; hard, HARD; soft nonthermal, SNTH; intermediate, INT; ultra soft, US; and very high, VH), the EUV fluxes, bolometric luminosities, and wind velocities at a compact star distance (without gravitational pull) are given in   Table~\ref{data}.
Examples of the wind simulations are shown in Fig.~\ref{loop}. 

To guarantee convergence, it is essential to use a reliable fitting  model. We used
 the $\beta$-velocity model ($V$ = $V_{inf}$(1 - $R_{star}$/$r$)$^{\beta}$) but multiplied it by  a suppression factor $SF$. In this way the model mimics the bowed (due to EUV irradiation) velocity curves of Fig.~\ref{loop} below the kinks, and $SF$ can be written as

\begin{equation}
SF = 1 - a\times(x - b)/(3.4 - b)
,\end{equation}

where $a$ and $b$ are the fitting parameters and $x$ = $r/R_{star}$ is the distance from the WR surface ( $x$ = 1 and  3.4 at the surface and CS distance, respectively). The fitting with the model was performed when the velocity gradient was large enough to provide a positive line force (left from  the kinks in the curves of Fig.~\ref{loop}). When the gradient became small, the force vanished (strongly depending on the velocity gradient), and the velocity and density values at the kink  determined the wind particle  flying outward. Vilhu et al. (2021) used another form of $SF$ above the kink, but we consider the present approach better. One could in principle combine these two suppression factors providing too many free parameters.

\begin{table}

\caption{EUV fluxes ($EUV$=$EF_E$ in units of 2$\times$$10^{36}$ erg/s at 100 eV), model luminosities ($L_{bol}$ =  $L_{eqpair}$ $10^{38}$ erg/s), and wind velocities ($V_{wind}$ km/s) without the gravitational pull of the compact star at its distance on the face-on side and wind-fed accretion power $P_{acc}$ ($10^{38}$ erg/s)  for the spectral states used. The acronyms used stand for the following: NOX=no X-ray irradiation, HARD=hard, SNTH=soft nonthermal, INT=intermediate, US=ultra soft, and VH=very high. The standard NOX $\beta$-velocity model has $V_{inf}$ = 1600 km/s and $\beta$ = 0.5. The corrected $V_{wind}$-values  are slightly different from those in the published A\&A-article Vilhu et al. (2023).}

\begin{tabular}{lcccccccc}

&&& &  & && \\
 PARAM&NOX &HARD&SNTH &INT &US&VH\\
$EUV$&0&0.25&0.45&0.77&1.04&1.26\\
$L_{eqpair}$&0&1.56&2.18&2.02&3.51&4.05\\
$V_{wind}$&1360&1010&906&840&797&762\\
$P_{acc}$&1.35&3.72&4.32&5.21&5.91&6.55\\

\end{tabular}

\label{data}
\end{table}

In the BHL formalism, the accretion radius is 

\begin{equation}
R_{acc}=2GM_{CS}/V_{rel}^2
,\end{equation}

where the relative velocity $V_{rel}$ was computed from the wind velocity at the compact star location and the orbital velocity of the compact star as
\begin{equation}
V_{rel}^2 = V_{wind}^2 + V_{orb}^2
,\end{equation}

where $V_{orb}$ = 675 km/s for the masses used. The gravitational pull of the compact star is not included because it is taken into account in the BHL formalism itself.

The wind-fed accretion power was then computed from 

\begin{equation}
P_{acc} = \dot M_{acc}\times \eta c^2
,\end{equation}
where the mass accretion rate is

\begin{equation}
\dot M_{acc} = \dot M_{wind}\times \pi R_{acc}^2/(4\pi A^2)
,\end{equation}

where $A$ is the binary separation (3.4$R_{\odot}$), $\dot M_{wind}$ is the mass-loss rate of the WR star,
and the mass-to-energy conversion factor $\eta$ = 0.1 is approximately valid for neutron stars and black holes (Frank, King and Raine 1992).

We assume that  a part of the wind-fed power $P_{acc}$ goes to the radiation luminosity $L_{rad}$ of the X-ray source, and a part of it  goes to advection $L_{adv}$ (negative) and kinetic energy of the jets and accretion disk wind  $L_{kin}$ (Narayan and Yi, 1994; notation from Fender and Munoz-Darias, 2016):

\begin{equation}
P_{acc} = L_{rad} + L_{kin} - L_{adv}
.\end{equation}
The bolometric radiation luminosity of the source $L_{eqpair}$ (derived from EQPAIR modeling) is the intrinsic property of the X-ray source and is equal  to $L_{rad,}$ 

\begin{equation}
L_{rad} = L_{eqpair} + \epsilon
,\end{equation}

 but  we assume a  small variability $\epsilon$  in this equality. Either  
 $L_{eqpair}$ or  $L_{rad}$ vary, or both of them do. When $L_{rad}$ is responsible for $\epsilon$ ($L_{eqpair}$ being constant), this means, for example, that  advection   changes with time, and the physics of which depends on the flow (see e.g., Esin et al. 1997, Liu et al. 2006b, Cao 2016). In this case, $L_{eqpair}$($EUV$) is a time average of $L_{rad}$($EUV$). When discussing masses, Vilhu et al. (2021) assumed that all wind-fed power goes to radiation. This, however, did not influence their wind modeling.

\begin{figure}
   \centering
   \includegraphics[width=9cm]{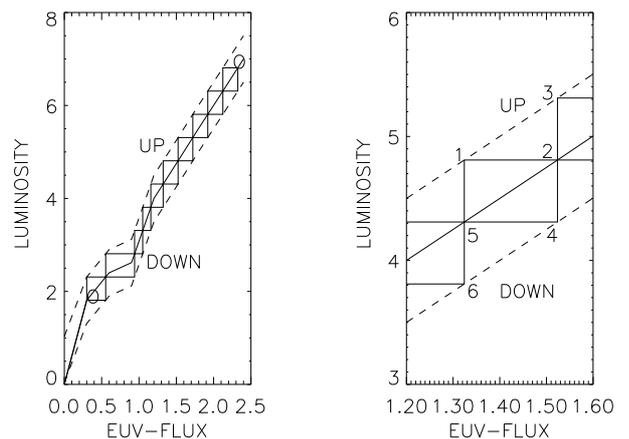}
     \caption{Basic idea of the present paper.  
Left panel: Bolometric X-ray source luminosity $L_{eqpair}$ (in units of $10^{38}$ erg/s) of Fig.~\ref{LvsEUV} is shown  versus the EUV irradiation from the X-ray source  ($EF_E$ in units of 2$\times$$10^{36}$ erg/s at 100 eV) (solid line). Around it, $L_{rad}$ is variable and two positions are shown (dashed lines). When $L_{rad}$ - $L_{eqpair}$ = $\epsilon$ is positive, the evolution goes upward, "UP" (1-2-3, positive feedback), when negative it goes downward, "DOWN" (4-5-6, negative feedback).  Right panel: Zoomed version of the panel on the left.}
         \label{zigzag1}
\end{figure}

Fig.~\ref{zigzag1} gives the basic idea of the present study, in the case when   $L_{eqpair}$($EUV$) is constant and   $L_{rad}$($EUV$) is varied by $\epsilon$(time). The results are identical if  $L_{eqpair}$ is variable. The only difference is that the evolution in Fig.~\ref{zigzag1} goes counterclockwise (UP and DOWN change place). The solid line is  $L_{eqpair}$ around which the dashed lines, representing  $L_{rad}$, vary at  the $\epsilon$ distance. In the plot two $\epsilon$ values are shown (+0.5 and -0.5).   There is a delay between EUV fluxes  from wind-fed power and that from the source. We modeled this  feedback  using the viscosity timescales at the circularization radius and timescales from the ASM light curve, and assuming different time profiles for $\epsilon$.  

At point 1 (Fig.~\ref{zigzag1}, right panel), the radiation luminosity$L_{rad}$ was determined from the wind with EUV flux there. The source  finds a state corresponding to this luminosity  which, in turn, determines the EUV flux (at point 2) and corresponding wind velocity and accretion luminosity (point 3). In this way the process continues upward (positive feedback). In the case in which $\epsilon$ is negative, the process goes downward (points 4-5-6, negative feedback). We assume that the output luminosity comes from  $L_{rad}$ and  $L_{eqpair}$ is its time average. A fraction of the wind-fed power $P_{acc}$ goes to kinetic energy and advection (see Fig.~\ref{LvsEUV}).

We set  the viscosity time at the circularization radius for the duration of one step in the upward and downward channels (1-2-3 or 4-5-6). One step corresponds, on average, to a luminosity change around $10^{37}$ erg/s. This choice determines how steep the rises and declines are.  In numerical simulations of Section 3, other choices are also presented. The time spent in high and low states is determined by the time profile of $\epsilon$. At EUV fluxes 0.25 and 1.26 (hard and very high states), these viscosity times are 1.1 and 4.1 days, respectively, as computed from 
the formula given in Cao and Zdziarski (2020):
\begin{equation}
 t_{visc} =  77(R/0.9RL)^{0.5}*10^5/T*\mu/(4/3)*(M_{CS}/5)^{2/3}*0.1/\alpha 
,\end{equation}
where  $t_{visc}$ is in days, $R$ is the distance from the compact star, and $RL$ is its Roche-lobe radius computed from the standard formula 
\begin{equation}
RL=A*0.46*(q/(1+q))^{1/3}
,\end{equation}
where $A$ is the binary separation and $q$ is the mass ratio $M_{CS}/M_{WR}$. For the masses used, $RL$ = 0.88$R_{\odot}$. We assume that $\mu$ =4/3 (the mean molecular weight of fully ionized Helium gas), $T$ = $10^5$ K (radiation black-body temperature of the WR star), and $\alpha$ = 0.1 (viscosity parameter). 

The circularization radius was computed from (see e.g., Kretschmar et al. 2021)
\begin{equation}
R_{circ}=(1+1/q)/4*A*(R_{acc}/A)^4
.\end{equation}

 The upward process continues until the accretion radius $R_{acc}$ is close to the Roche-lobe radius $RL$ or else the increasing EUV flux combined with the decrease in clumping stops the wind, and the process starts again. In Fig.~\ref{zigzag1} (left), the maximum value  is marked (see the oval "O" in the upper right corner).   One can also speculate that changing the filling factor from 0.1 (the baseline value) to 0.3 may lead to the cessation of the wind during the very high state (see Fig.~\ref{loop}). This influences the highest luminosity reached.

In our simulations (next section), we used the 0.9$\times$ $RL$ limit for the maximum accretion radius, which gave  a maximum luminosity of 5.3$\times$ $10^{38}$ erg/s (a bit less than the Eddington luminosity for ionized helium). This limit was a random choice, but we followed the reasoning from  Cao and Zdziarski (2020) that the accretion disk is tidally limited by 0.9 $\times$$RL$. We assumed the same for the accretion radius.  The system stays at a  high level unless the channel down opens ($\epsilon$ becomes negative, starting negative feedback). The oval "O" in the lower-left corner of Fig.~\ref{zigzag1} (left) shows the lowest level where the system ends after the downward motion. In our simulations it was set to the NOX accretion power value  1.35$\times$$10^{38}$ erg/s . When the upward channel opens ($\epsilon$ becomes positive, positive feedback), a new climb upward starts.
In the next section, we illustrate this process by model computations and propose the origin for the  ASM light-curve variability.

\section{Numerical results}
We show numerical simulations based on the previous section, assuming that $L_{rad}$ is variable using two types of time profiles for $\epsilon$ (in Eq. 7): i) a random box-style profile and ii) a profile that mimics the ASM daily light curve.

\begin{figure}
   \centering
   \includegraphics[width=9cm]{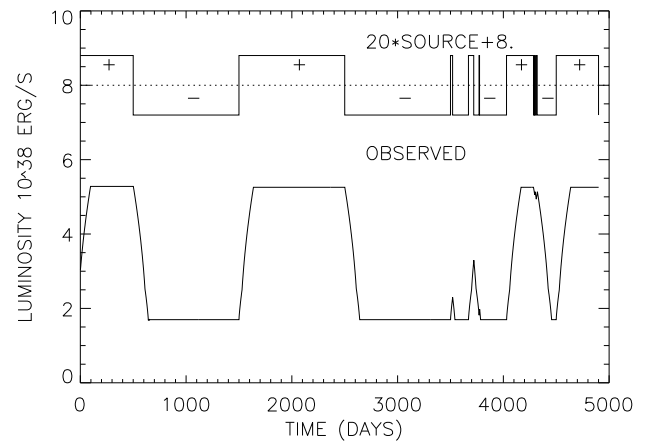}
     \caption{Upper part of the plot: Randomly selected box-like time profile for $\epsilon$  (=SOURCE, in the plot multiplied by 20). Lower plot: Observed luminosity after the feedback steps outlined in Section 2. The plus and minus signs in the upper plot mark the times of positive and negative feedback, respectively ($\epsilon$ positive or negative).   See text for details.}
         \label{lc_boxes}
\end{figure}

\begin{figure}
   \centering
   \includegraphics[width=9cm]{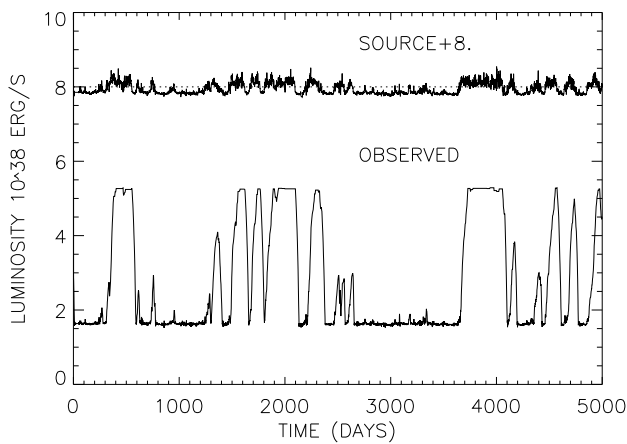}
     \caption{Upper plot: Assumed $\epsilon$ profile in Eq. 7 resembling the ASM daily light curve (=SOURCE). Lower plot: Observed luminosity after the feedback process outlined in Section 2.  See text for details.}
         \label{LC_asm}
\end{figure}

\begin{figure}
   \centering
   \includegraphics[width=9cm]{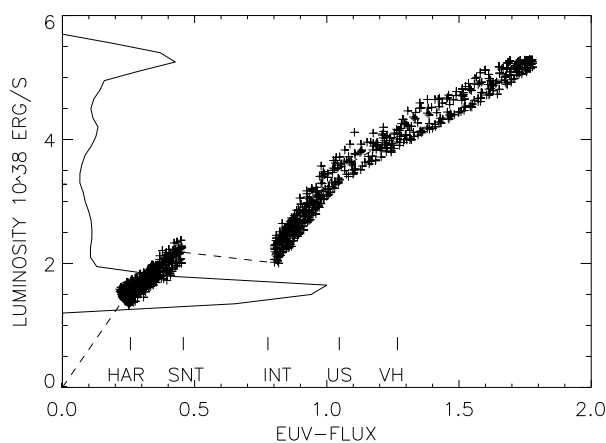}
     \caption{ Luminosity (in units of $10^{38}$ erg/s) plotted against the EUV flux ($EF_E$ in units of 2$\times$$10^{36}$ erg/s at 100 eV) for the simulation of Section 3.2 shown in Fig.~\ref{LC_asm}. The mean location of spectral states are marked as follows: HAR=hard, SNT=soft nonthermal, INT=intermediate, US=ultrasoft, and VH=very high.  The histogram on the left (thick solid line) shows the relative number of days spent at fixed luminosity.  The dashed line shows $L_{eqpair}$.
 }
         \label{LEUVLOOP}
 \end{figure}

\begin{figure}
   \centering
 \includegraphics[width=9cm, height=10cm]{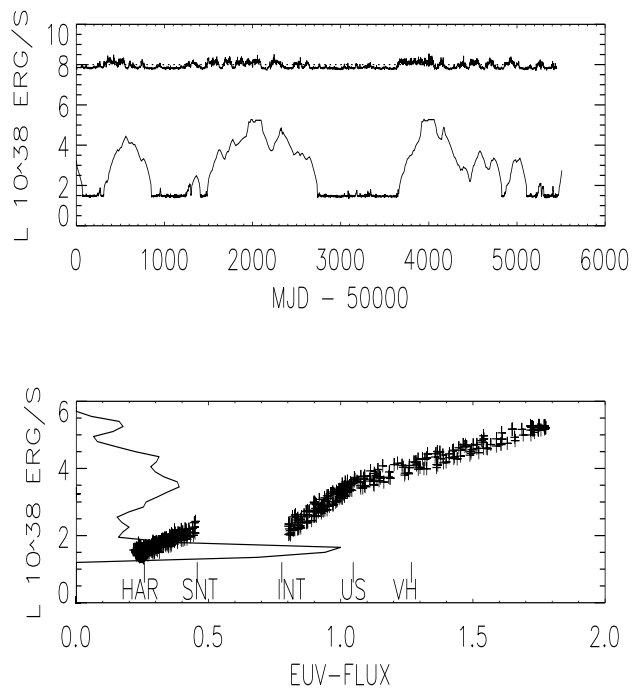}

     \caption{Same as Fig.~\ref{LC_asm} (upper panel) and Fig.~\ref{LEUVLOOP} (lower panel), but with 4 times longer rise and decline times in the light curve. }
         \label{ANAL6}
\end{figure}

\subsection{Box profile}
We made a random box-like $\epsilon$ profile, which is shown in the upper part of Fig.~\ref{lc_boxes}. Boxes were selected  to show the effect of different timescales during low and high states. The profile is symmetrical around zero mean. The amplitude of the profile is not very constraining and  could have been either smaller or larger, resulting in a similar observed luminosity after the feedback steps (Fig.~\ref{lc_boxes} bottom).  The rise and decline parts are slightly tilted due to the viscosity timescales used (1.1 and 4.1 days at hard and very high state levels, respectively) for one step  with ann average luminosity change of $10^{37}$ erg/s. Timescales for other phases were linearly interpolated from these two values. 
\subsection{ASM profile}
In the simulation, the time profile of $\epsilon$ was taken directly from the observed ASM daily  light curve by $\epsilon$ = 0.02$\times$(ASM counts/s -15; the upper part of Fig.~\ref{LC_asm}). The normalization of $\epsilon$ is ad hoc. However, the results are not very sensitive to the amplitude of $\epsilon$. The profile is rather symmetrical around zero, but shifting the zero level lower would not change the results significantly. The  essential property of the $\epsilon$ time profile are  turnovers of its sign (changing to positive or negative feedback). 

Using  the selected expression for  $\epsilon$ basically means picking  turnover times from the ASM daily light curve at 15 counts/s level. The rise and decline times used for one step  with an average luminosity change of $10^{37}$ erg/s were the viscosity timescales at circularization radii (1.1 and 4.1 days at hard and very high state levels, respectively). 
The lower part of  Fig.~\ref{LC_asm} shows the resulting light curve  explained in Section 2. As expected, this light curve is qualitatively very similar to the $\epsilon$ time profile (upper plot) reflecting its turnover times. 

 Fig.~\ref{LEUVLOOP} shows the evolution of the light curve in the $EUV$-$L$ diagram.
Luminosity (in units of $10^{38}$ erg/s) is plotted against the EUV flux ($EF_E$ in units of 2$\times$$10^{36}$ erg/s at 100 eV) for the simulation shown in Fig.~\ref{LC_asm}. The mean location of spectral states are marked.   The dashed line shows $L_{eqpair}$. The histogram on the left (thick solid line) shows the relative number of time spent at fixed luminosity.

Fig.~\ref{ANAL6} shows the effect of lengthening the rise and decline times by a factor of 4  from those used in Figs.~\ref{LC_asm} and ~\ref{LEUVLOOP}. This smooths the light curve and the  luminosity histogram.

\section{Observed ASM  histogram}

In the previous section, luminosity histograms were shown for the simulated light curves (see Figs.~\ref{LEUVLOOP} and ~\ref{ANAL6}, lower plot). We present a similar histogram for the ASM light curve to be compared.

\begin{figure}
   \centering
   \includegraphics[width=9cm]{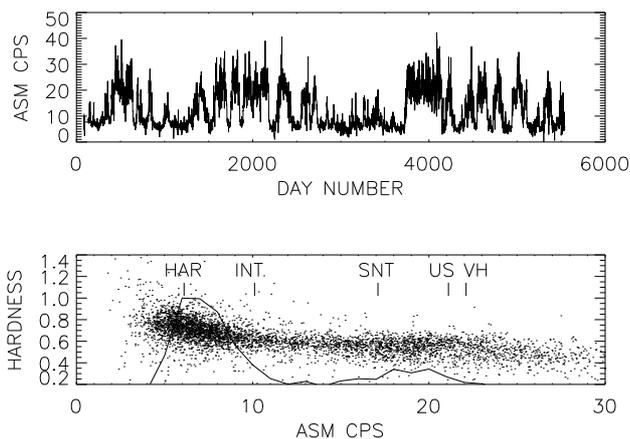}
     \caption{Upper panel: $RXTE$/ASM daily light curve of Cyg X-3. Lower panel: Hardness [C-A]/[C+A] versus count rate (dots). The solid line shows the histogram (relative number of days/ASM bin). The mean positions of spectral states are marked as follows: HAR=hard, SNT=soft nonthermal, INT=intermediate, US=ultrasoft, and VH=very high.
 }
         \label{MULTIPLOT}
 \end{figure}

 A  partial gap between low and high states  exists  in the observed ASM light curve, which is illustrated in Fig.~\ref{MULTIPLOT} (lower panel), where the hardness [(C-A)/(C+A)] is plotted against the daily mean count rate. The energy range of ASM is 1.5 - 12.1 keV with  filters A (1.5 - 3 keV) and C (5 - 12.1 keV). The solid line in the lower plot shows the histogram of the relative number of days spent in the ASM daily count rate bin (1 ct/s). Mean count rates of the spectral states are marked as estimated from the observing log of Hjalmarsdotter et al. (2009). We note that the places of the intermediate and soft nonthermal states are reversed in Figs.~\ref{LEUVLOOP} and ~\ref{MULTIPLOT}. This is due to their different spectral shapes (see Fig.~\ref{spectralstates}). 

Comparing the simulated histograms of  Section 3.2 with the observed ASM   histogram  (Fig.~\ref{MULTIPLOT}, lower panel), we suggest a variable maximum accretion radius (in simulations set to 0.9$\times$$RL$)  and/or variable timescales of  rise and decline phases of the  light curve. Both of these smear the histogram closer to that of ASM. 
%

\section{Timescales from the ASM light curve}

The timescale estimate used during the rising and declining parts of the light curve was crude (1.1 and 4.1 days during the hard and very high state, respectively). We compare it with that derived from the ASM light curve itself. 

\begin{figure}
   \centering
   \includegraphics[width=9cm]{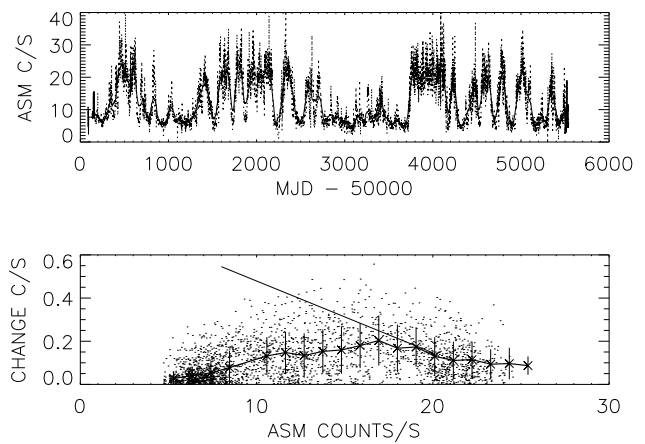}
     \caption{Upper panel: $RXTE$/ASM daily light curve of Cyg X-3 (the dotted line) overplotted with the smoothed version used for timescale estimates (the solid thick line). Lower panel: Mean daily changes (counts/sec, the solid thick line) with  error bars. Dots give individual daily changes. The solid linear line gives a comparison with the timescales used in Section 3 (see the text).
 }
         \label{ASM}
 \end{figure}

We built a smoothed version of the daily ASM  light curve shown in  Fig.~\ref{ASM} (upper panel, solid line). The original light curve is shown by the dotted  line.  The smoothing was done by a box car with a 60 days width. From this smoothed light curve, mean differences (absolute values) between subsequent days were determined at binned ASM count rate intervals and their standard deviations were computed. This is shown in  Fig.~\ref{ASM} (lower panel). The dots show individual changes during one day, while the solid thick line gives the mean value with one-sigma errors.

Assuming  that the luminosity $L$ (in units of $10^{38}$ erg/s) can be approximated by  the ASM count rate divided by six, and that the feedback step is around  $10^{37}$ erg/s, we can mark the timescales of our simulations in Fig.~\ref{ASM} (lower panel, the solid linear line). The approximate factor 6 comes from comparisons of ASM count rates with EQPAIR luminosities.

\begin{figure}
   \centering
   \includegraphics[width=9cm,height=10cm]{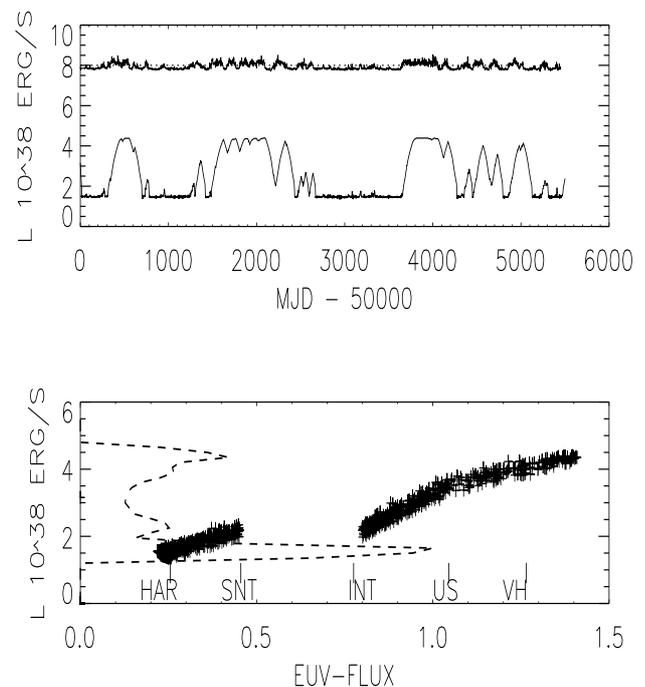}
     \caption{ Results of simulations with  timescale estimates from smoothed ASM light curve (see Fig.~\ref{ASM} and text). Upper panel: Same as Fig.~\ref{LC_asm}. Lower panel: Same as Fig.~\ref{LEUVLOOP}.  
 }
         \label{ANAL2}
 \end{figure}

The timescale for the feedback steps  can be computed as follows.
Marking the mean value of the daily change by YASM (Fig.~\ref{ASM}  lower panel, thick solid line with error bars),  then the time spent for one step is DL/(YASM/6) where DL is the luminosity change during one step (in units of $10^{38}$ erg/s). Using this timescale,  we  repeated the computations performed in Section 3. The results are shown in Fig.~\ref{ANAL2}. These results are between those  in Section 3.2 using $t_{visc}$ and 4$t_{visc}$. Due to the scaling used (the factor 6), the maximum luminosity level is also lower. 

\section{Discussion}

Without EUV irradiation of the X-ray source, it is impossible to provide enough power for all spectral states (see Fig.~\ref{LvsEUV}), provided that the mass of the compact star and mass-loss rate of the WR star are not higher than what is  assumed in the present paper (2.4$M_{\odot}$  and  0.65$\times$10$^{-5}$  $M_{\odot}$/year).
As an example, the combination of 3.5$M_{\odot}$  and  1.0$\times$10$^{-5}$  $M_{\odot}$/year would give enough accretion power for all spectral states by the pure wind model without EUV irradiation (see Fig.~\ref{LvsEUV}, the horizontal NOX line would rise threefold). These parameter values are within acceptable ranges. However, even in this case, the wind-fed power has a positive correlation with the EUV flux and the process outlined here should work.

 We have proposed  a heuristic feedback model
 to relate the X-ray variability of Cyg X-3 to the X-ray irradiation on the face-on side of the WR companion, and subsequent lowering of the wind velocity and increasing  wind-fed accretion. 
The Bondi-Lyttleton-Hoyle modeling used to estimate the wind-fed power (Eq. 4) is approximate and very sensitive to the wind velocity. For our purposes, the plot in Fig.~\ref{LvsEUV} 
demonstrates that the wind-fed power is larger than what the source radiates. In the present work, we have assumed that a part of it goes to radiation ($L_{rad}$), which is varying by a small factor  $\epsilon$ (eq. 7).  In the  model a small seed $\epsilon$ can produce the ASM light curve, although the process  works for different amplitudes of  $\epsilon$.  The size of  $\epsilon$ is not crucial with the essential property being turnover times when $\epsilon$ changes its sign and starts positive or negative feedback.

The kink in $L_{eqpair}$ around EUV flux = 0.7 (see Figs.~\ref{LvsEUV} and ~\ref{LEUVLOOP}) may have some meaning, but it is not crucial here because identical results can be observed by straightening the kink. The kink is caused by the nonlinear behavior of  $L_{eqpair}$ around the soft nonthermal and intermediate states (see Fig.~\ref{spectralstates}) and should be confirmed.
The main reason for  the kink seems to be the intermediate state spectral model (Fig.~\ref{spectralstates}). The intermediate state refers to a transitional spectral state between the hard and soft states.  The low ASM  count rate of the intermediate state 
 (Fig. \ref{MULTIPLOT}) supports the kink and low bolometric luminosity.

We approximated the physics involved by an ad hoc scaling law type process. The ends of the high and low states (ovals "O" in Fig.~\ref{zigzag1}) were set by the Roche-lobe and NOX limits, respectively. Their durations, in turn, are due to opening of the upward and downward channels (turnover times of $\epsilon$).  We have assumed that this process takes place in steps between the baseline $L_{eqpair}$, characterizing the X-ray source, and a small deviation from it ($L_{eqpair}$ $\pm$ $\epsilon$). The real physics of this process remains to be clarified, but first steps could be done as follows. A simple case would be when the advective and kinetic parts in eq. 6 were zero: $P_{acc}$ = $L_{rad}$. Then the feedback process could be done without injecting $\epsilon$ into eq. 7. If the X-ray source model was simple as in an accretion disk, one should be able to directly follow the feedback process without $\epsilon$, provided one has access to a physical time-dependent accretion disk code. Further, if the advective and kinetic parts were not zero, one could repeat this  as outlined in the present paper and compare the results. 

In each upward and downward step, the EUV flux changes, affecting the wind velocity and mass accretion rate.   We assumed that the wind relaxation takes place more rapidly (in less than hours) than the viscosity timescale of the disk at a circularization radius (days).  Wind particles travel across the wind in less than half an hour (Vilhu et al. 2021) and thermal timescales are very small because the wind mass is small (on the order of $10^{-9}$$M_{\odot}$) .

 Variations in  $\epsilon$ can  be caused by variable kinetic energy (jets, accretion disk wind) and/or advection,  the physics of which are not fully understood (see e.g., Esin et al. 1997, Liu et al. 2006b, Cao 2016).  
To explain the smooth distribution of ASM count rates (the histogram in Fig.~\ref{MULTIPLOT}, lower panel), one has to assume either a variable maximum accretion radius  and/or changing timescales during the rise and decline parts of the light curve. The effects of these are illustrated by histograms in Figs.~\ref{LEUVLOOP}, ~\ref{ANAL6}, and ~\ref{ANAL2}, and clearly need further work.

Many  wind-fed X-ray binaries are expected to experience a qualitatively similar behavior, for example Cyg X-1, which is a very tight system with a binary separation about twice the OB-supergiant radius. For comparison,  in Cyg X-3 the separation is 3.4 times the WR-star radius.

\section{Conclusions}
We  propose a simple heuristic model to investigate the feedback between the WR wind and the X-ray source of Cyg X-3. In our model, the radiation from accretion $L_{rad}$ fluctuates around the mean value $L_{eqpair}$  characteristic of the X-ray source and derived from modeling time-averaged X-ray spectra. There is a delay between EUV fluxes  determining the wind-fed power and the bolometric luminosity. 

We modeled this  feedback using viscosity timescales   and assumed different time profiles for variability ($\epsilon$ in Eq. 7). More realistic timescales were estimated from the smoothed ASM light curve (Section 5). These timescales determine the 
rise and decline times of the  light curve, while the $\epsilon$ time profile sets limits on the  duration of low and high states. The essential property of   $\epsilon$ are turnover times when it changes sign and starts positive or negative feedback.

We  have demonstrated how  light-curve variations of Cyg X-3 can have their origin in small temporal changes  of the accretion flow  ($\epsilon$ in Eq. 7). These changes may be variations in advection and/or kinetic energy (jets, accretion disk wind).
The time profile of the resulting luminosity mimics the input $\epsilon$ but it is magnified (see Figs.~\ref{lc_boxes} and ~\ref{LC_asm}). Comparing the histograms of Figs~\ref{LEUVLOOP}, ~\ref{ANAL6}, ~\ref{MULTIPLOT}, and ~\ref{ANAL2},  additional variable parameters (besides $\epsilon$) are needed. We suggest a variable maximum accretion radius and/or changing timescales for the rise and decline parts of the light curve.

\begin{acknowledgements}
This project has received funding from the European Research Council (ERC) under the European Union's Horizon 2020 research and innovation programme (grant agreement No. 101002352). We thank Dr Linnea Hjalmarsdotter for permission to use her Fig. 1.
\end{acknowledgements}


\end{document}